# The discrete analogue of the method of quickest descent for an inverse acoustic problem in case of a smooth source

G. Tyulepberdinova, G.Gaziz, N. Kerimbayev, S. Abdykarimova
al-Farabi Kazakh National university, Almaty, Kazakhstan

**Abstract**

The article considers the discrete analogue of the method of quickest descent for an inverse Acoustics problem in case of a smooth source. The authors derived the gradient of functional in differential and discrete cases, described the algorithm of solving a problem, and compared gradients of functional in continuous and discrete cases. In the article the improved estimates of the rates of convergence of gradient-based methods are obtained, which are very important for practice because they provide with the possibility to make input data errors consistent with the iteration number. There is a practical application of the proposed new method of deriving the gradient of functional for an Acoustics discrete problem, for it provides with calculations that are more accurate. The theoretical importance of the method is the developed technique of deriving estimates and gradients of functional at a discrete level.

## 1 Introduction

In Acoustics, inverse problems are understood as restoration of sound sources or heterogeneousness (inhomogeneous) characteristics, which scatter a primary field on the basis of measuring the primary or scattered acoustic field. Historically, first solving methods used to be based on approximation of a single scattering and smooth variation of scattering characteristics. However, the assumption used in these approximations seriously constrain the field of their application. Further investigations connected with taking into account effects of multiple scatterings, have shown that the inverse scattering problem is incorrect and nonlinear in unknown functions. One of methods of solving inverse problems, which consider multiple scatterings, is the iteration method.

The objective of this work is a proper study and derivation of the gradient of functional in the differential and discrete analogues using the method of quickest descent for an inverse acoustics problem in case of a smooth source.

A.L. Milman dedicated his investigations in the sphere of scattering theory for a wave equation in two-dimensional space. The paper "The Inverse Scattering Problem for Acoustic Centro Symmetrical Finite Objects in Two-dimensional Space", presents Scattering theory for a wave equation in two-dimensional space perturbed by a finite function of a radial variable. There has been shown that this compression has one-dimensional defect subspaces, and its characteristic operator-function is a meromorphic function (1990).

J.R. McLaughlin, P.E. Sacks, M. Somasundaram in their publication "Inverse Scattering in Acoustic Media Using Interior Transmission Eigenvalues" consider well-known inverse problems for acoustic media consisting in the determination of the sound speed of the medium using far field data. A precise formulation is as follows. The paper considers the Helmholtz equation, where the sound speed, $c$ (z), satisfies some condition for some constant $c_0 > 0$ (1997).

Works by A.S. Blagoveschenskiy and D.A. Fedorenko are dedicated to the inverse problem for acoustics equations in the weekly horizontally inhomogeneous medium (2008).

The paper by V.A. Barkhatov, "Solution of the One-dimensional Inverse Acoustic Problem with Allowance for Velocity Dispersion and Frequency-dependent Wave Attenuation", considers a one-dimensional inverse acoustics problem. The solution of this inverse problem determines the dependence of the reflecting coefficient of echo-signals, which a defectoscope receives and processes at a distance. The solution of a one-dimensional inverse acoustics problem is considered taking into account velocity dispersion and attenuation dependent on wave frequency. Besides, the paper presents the results of numerical experiments(2009).

## 2. Formulation of the problem
Let us consider the one-dimensional acoustic equation:

$$\frac{1}{\rho c^2} w_{tt}(z,t) = \left(\frac{1}{\rho} w_z(z,t)\right)_z, \quad z > 0, \quad t > 0 \tag{1}$$

Here $c(z) > 0$ is the sound speed, $p(z) > 0$ is the density, $\omega(z,t)$ is the pressure. Equation (1) describes the propagation of acoustic waves of a small amplitude in a subspace satisfying the following conditions

$$w(z,t)|_{t<0} = 0, \tag{2}$$

$$w_z(+0,t) = \theta(t), \quad t > 0, \tag{3}$$

$$w(+0,t) = f(t), \quad t > 0, \tag{4}$$

where $\theta(t)$ is a smooth function. It is well known that using an additional information in the form (4), it is difficult to determine the two functions, $c(z)$ and $p(z)$, simultaneously. Therefore, we formulate the inverse problem of determining one function. For this purpose, we introduce a new variable and function.

$$x = \varphi(z) = \int_0^z \frac{d\xi}{c(\xi)}, \quad z = \varphi^{-1}(x)$$

$$v(x,t) = w(z,t) = w(\varphi^{-1}(x), t)$$

$$h(x) = \rho(z) = \rho(\varphi^{-1}(x)), \quad g(x) = c(z) = c(\varphi^{-1}(x)).$$

Then (1) – (4) will have the forms

$$v_{tt}(x,t) = v_{xx}(x,t) - \frac{\sigma'(x)}{\sigma(x)} v_x(x,t), \quad x > 0, t > 0, \tag{5}$$

$$v(x,t)|_{t<0} = 0, \tag{6}$$

$$v_x(+0,t) = \theta, \quad t > 0, \tag{7}$$

$$v(+0,t) = f(t), \quad t > 0. \tag{8}$$

where $\sigma = g(x)h(x)$ is the acoustic impendence, $\alpha = \beta c(+0)$.

Further (V.G. Romanov), we reduce equation (5) to the following form:

$$u_{tt}(x,t) = u_{xx}(x,t) - q(x)u(x,t), \tag{9}$$

where

$$u(x,t) = v(x,t)\exp\left\{-\frac{1}{2}\ln\sigma(x)\right\}, \tag{10}$$

$$q(x) = -\frac{1}{2}[\ln\sigma(x)]'' + \frac{1}{4}\left[\frac{\sigma(x)}{\sigma(x)}\right]^2 \tag{11}$$

Further, in order to make investigation of the inverse problem at a discrete level more convenient, taking into account the above reduction, we consider the following problem. We will determine q(x) using the following expressions

$$\tilde{u}_{tt}(x,t) = \tilde{u}_{xx}(x,t) - q(x)\tilde{u}(x,t), \quad x \in R, \quad t > 0 \tag{12}$$

$$\tilde{u}(x,0) = \varphi_0(x), \quad \tilde{u}_t(x,0) = \varphi_1(x), \quad x \in R, \tag{13}$$

$$\tilde{u}(0,t;q) = \varphi_2(t), \quad \tilde{u}_t(0,t) = 0, \quad t > 0. \tag{14}$$

Let us consider the equivalent inverse problem (S.I. Kabanikhin, K.T. Iskakov):

$$u_{tt}(x,t) = u_{xx}(x,t) - q(x)u(x,t), \quad x \in R, \quad t > 0 \tag{15}$$

$$u(x,0) = \varphi_0''(x) - q(x)\varphi_0(x), \quad u_t(x,0) = 0, \quad x \in R, \tag{16}$$

$$u(0,t;q) = f(t), \quad u_x(0,t) = 0, \quad t > 0. \tag{17}$$

The expressions (15) – (17) follow from (12) – (14), if we substitute

$$u = \tilde{u}_{tt}, \quad f(t) = \varphi_2''(t), \quad \varphi_1(t) = 0.$$

When using the optimization method for solving an inverse problem, a criterion functional is minimized, for example a criterion functional of the following form:

$$J(p) = \int_0^T [u(0,t;p) - f(t)]^2 dt. \tag{18}$$

The notation u(0,t;p) indicates that u(x,t) is a solution of the direct problem (15) – (17) at fixed p(x). The minimization of the functional J(p) is realized by the method of quickest descent:

$$p^{(n+1)}(x) = p^{(n)}(x) - \alpha_n \nabla J[p^{(n)}],$$

in which the descent factor, $\alpha_n$, is determined using the condition

$$J(p^{(n)} - \alpha_n \nabla J(p^{(n)})) = \inf_{\alpha>0}\{J(p^{(n)} - \alpha_n \nabla J(p^{(n)}))\}, \quad \alpha_n > 0.$$

Here $\nabla J(p^{(n)})$ is the gradient of functional, n is the number of iteration.

### 3. Optimization method at differential level

Let us formulate the inverse problem under study concretely. In the domain $D = [0, L] \times [0, T]$ it is required to determine the function q(x) using the expressions

$$u_{tt} = u_{xx} - q(x)u, \ (x,t) \in D \tag{20}$$

$$\begin{cases} u(x,0) = q(x), \\ u_t(x,0) = 0 \end{cases} \tag{21}$$

$$\begin{cases} u_x(0,t) = 0, \\ u(L,t) = 0 \end{cases} \tag{22}$$

and the given additional information of the form

$$u(0,t;q) = f(t) \tag{23}$$

For illustrative purposes, we choose a rectangular domain D.

It is easy to see that the expressions (20) – (22) can be easily obtained from (15) assuming that $\varphi_0(x) = -1$.

Let $p(x)$ be the approximate solution of the inverse problem. As before, we consider the discrepancy functional

$$J(p) = \int_0^T [u(0,t;p) - f(t)]^2 dt \tag{24}$$

The essence of the optimization method is the following: we define the initial approximation $p^{(0)}(x)$, we determine the successive approximations using (19).

Let us get down to deriving the formula of the gradient for the functional (24) using the method analogous to those in the works (S.I.Kabanikhin and others, G.A.Tyulepberdinova, A.E.Yerdeneyeva).

Let us consider the increment $p(x) + \delta p(x)$ и $\delta u = u(x,t; p + \delta p) - u(x,t; p)$, then for the increment $\delta u(x,t)$, neglecting second-order small quantities, we get the following problem (A.T.Nurseitova, D.B.Nurseitov, G.A.Tyulepberdinova):

$$\delta u_{tt} = \delta u_{xx} - u\delta p - \delta p u, \ (x) \in D \tag{25}$$

$$\delta u(x,0) = \delta p, \quad \delta u_t = 0, \tag{26}$$

$$\delta u_x(0,t) = 0, \quad \delta u(L,t) = 0, \tag{27}$$

Now we multiply both parts of the equation (25) by the function ψ(x,t), and integrating over the domain D, we obtain

$$\int_0^T\int_0^L \psi(x,t)\delta u_{tt}\,dxdt = \int_0^T\int_0^L \psi(x,t)\delta u_{xx}\,dxdt$$

$$-\int_0^T\int_0^L \psi(x,t)u(x,t)\delta p(x)\,dxdt - \int_0^T\int_0^L p(x)\delta u(x,t)\psi(x,t)\,dxdt \qquad (28)$$

Denoting by $S_1, S_2, S_3, S_4$ each item of the latter expression and applying integration by parts to $S_1, S_2$, we obtain:

$$S_1 = \int_0^T\int_0^L \psi(x,t)\delta u_{tt}\,dxdt = \int_0^L\left[\psi\delta u_t\Big|_0^T - \int_0^T \psi_t(x,t)\delta u_t\,dt\right]dx$$

$$= \int_0^L\left[\psi\delta u_t\Big|_0^T - \left(\psi_t\delta u\Big|_0^T - \int_0^T \psi_{tt}(x,t)\delta u(x,t)\,dt\right)\right]dt$$

$$= \int_0^L [\psi(x,T)\delta u_t(x,T) - \psi(x,0)\delta u_t(x,0) - \psi_t(x,T)\delta u(x,T)$$

$$+ \psi_t(x,0)\delta u(x.0) + \int_0^T \psi_{tt}(x,t)\delta u(x,t)\,dt\Big]dx$$

Using the condition (26) and assuming that $\psi(x,T)=0,\ \psi_t(x,T)=0,$ we have

$$S_1 = \int_0^T\int_0^L \psi_{tt}(x,t)\delta u(x,t)\,dxdt + \int_0^T \psi_t(x,0)\delta p\,dx \qquad (29)$$

Let us reduce

$$S_2 = \int_0^T\int_0^L \psi(x,t)\delta u_{xx}\,dxdt = \int_0^T\left[\psi\delta u_x\Big|_0^L - \int_0^L \psi_x(x,t)\delta u_x\,dx\right]dt$$

$$= \int_0^T \left[ \psi \delta u_x \big|_0^L - \left( \psi_x \delta u \big|_0^L - \int_0^L \psi_{xx}(x,t) \delta u(x,t) dx \right) \right] dt$$

$$= \int_0^T [\psi(L,t) \delta u_x(L,t) - \psi(0,t) \delta u_x(0,t) - \psi_x(L,t) \delta u(L,t)$$

$$+ \psi_x(0,t) \delta u(0,t) + \int_0^L \psi_{xx}(x,t) \delta u(x,t) dx ] dt$$

Using the condition (27) and assuming that $\psi(L,t) = 0$, $\psi_x(0,t) = 2[u(0,t;p) - f(t)]$, we obtain

$$S_2 = \int_0^T \int_0^L \psi_{xx}(x,t) \delta u(x,t) dt dx + \int_0^T 2[u(0,t;p) - f(t)] \delta u(0,t) dt \qquad (30)$$

Further, we leave the formulae for $S_3, S_4$ alone, that is

$$S_3 + S_4 = -\int_0^T \int_0^L \psi(x,t) u(x,t) \delta p(x) dx dt -$$

$$- \int_0^T \int_0^L p(x) \delta u(x,t) \psi(x,t) dx dt$$

Using (29) and (30) the formula (28) will have the form

$$\int_0^T \int_0^L \psi_{tt}(x,t) \delta u(x,t) dt dx = \int_0^T \int_0^L \psi_{xx}(x,t) \delta u(x,t) dx dt +$$

$$+ \int_0^T 2[u(0,t;p) - f(t)] \delta u(0,t) dt - \int_0^T \int_0^L \psi(x,t) \delta u(x,t) p(x) dx dt -$$

$$- \int_0^T \int_0^L \psi(x,t) u(x,t) \delta p(x) dx dt$$

Assume that $\psi_{tt} = \psi_{xx} - p(x)\psi$, then in the latter expression we have

$$\int_0^T 2[u(0,t;p)-f(t)]\delta u(0,t)dt = \int_0^T\int_0^L \psi(x,t)u(x,t)\delta p(x)dtdx.$$

The left part is equal to the increment of functional $\Delta J(p)=J(p+\delta p)-J(p)$, and the right part is its gradient. Therefore, the gradient of functional is equal to

$$\nabla J(p) = \int_0^T \psi(x,t)u(x,t)dt \tag{31}$$

Thus, in the process of deriving the formula for finding the gradient of a functional, we get an auxiliary dual problem for $\psi(x,t)$, which has the form

$$\psi_{tt} = \psi_{xx} - p(x)\psi, \qquad (x,t)\in D \tag{32}$$

$$\psi(x,T)=0, \quad \psi_t(x,T)=0, \tag{33}$$

$$\psi_x(0,t)=2[u(0,t;p)-f(t)], \tag{34}$$

$$\psi(L,t)=0 \tag{35}$$

Let us introduce the general scheme of the optimization method at a differential level (A.T.Nurseitova, D.B.Nurseitov, G.A.Tyulepberdinova):

1  Define the initial approximation $p^{(0)}(x)$, solve the direct problem (20) – (22) assuming that $q(x)=p^{(0)}(x)$, and find $u^{(0)}(x,t;p^{(0)}(x))$.

2  Calculate the value of the functional (24), if it attained the minimum then take $p^{(0)}(x)$ as an approximate solution of the inverse problem, if it did not then go to the next step.

3  Calculate the boundary value condition (34) when $p(x)=p^{(0)}(x)$, and solving the problem (32) – (35), we obtain its solution $\psi^{(0)}(x,t;p^{(0)}(x))$.

4  Find the gradient of the functional using the formula (31).

5  Using the formula (19) we find the next approximation $p^{(1)}(x)$.

6  Again calculate the value of the functional (24), if it attained the minimum then we take $p(x)=p^{(\lambda)}(x)$ as an approximate solution, if it did not then assuming $p^{(0)}(x)=p^{(\lambda)}(x)$ we return to step 3.

**4. Discrete analogue of the optimization method**

Let $p(x_i)$ be an approximate solution of the inverse problem. We approximate the problem (20) – (22) using the following difference scheme:

$$y_{\bar{t}t} = \hat{y}_{\bar{x}x} - p_i\hat{y}, \quad (x_i,t_j)\in w_{h\tau}, \tag{36}$$

$$y_i^0 = p_i, \quad y_{t.i}^0 = 0, \quad i=0,1,2,\ldots,N, \tag{37}$$

$$y_{x,0}^j = 0, \quad y_N^j = 0, \quad j=2,3,\ldots,M. \tag{38}$$

Here $\overline{w}_{h\tau} = \overline{w}_h \times \overline{w}_\tau$ is a net analogue of the domain D = [0, L] x [0, T],

$$\overline{w}_h = \{x_i = ih, \quad i = 0,1,2,\ldots,N, \quad h = L/N\},$$
$$\overline{w}_\tau = \{t_j = j\tau, \quad j = 0,1,2\ldots,M, \quad \tau = T/M\}.$$

$y_i^j = y(x_i, t_j)$, $\hat{y} = y_i^{j+1} = y(x_i, t_{j+1})$ are net approximations of the function u(x,t).

Let us assume that for the difference direct problem (36) – (38) there is an additional information. Let

$$y_0^j = f(t_j), \quad j = 2, 3, \ldots, M. \tag{39}$$

Let us consider the discrete analogue of the functional (24) of the form

$$J[p] = \tau \sum_{j=1}^{M} [y_0^j\{p_i\} - f^j]^2. \tag{40}$$

Let us define the increment $p_i + \delta p_i$, and $\delta y_i = y_i^j\{p_i + \delta p_i\} - y_i^j\{p_i\}$. further we will use the designation: $y_i^j = \{p_i\}$, which shows a particular dependence of the net function $y_i^j$ on the desired coefficient $p_i$. It is not difficult to obtain the following difference problem subject to the increment $\delta y_i^j$:

$$\delta y_{\bar{t}t} = \delta \tilde{y}_{\bar{x}x} - \hat{y}\delta p - p\delta\tilde{y}, \quad (x_i, t_j) \in w_{h\tau}, \tag{41}$$

$$\delta y_i^0 = \delta p, \quad \delta y_{t,i}^0 = 0, \tag{42}$$

$$\delta y_{x,i}^j = 0, \quad \delta y_N^j = 0. \tag{43}$$

Let us multiply both parts of the difference expression (41) by the net function $\varphi_i^j$ and summing up over j from 1 to M – 1 and over the index $i$ from 1 to N, we obtain:

$$\sum_{j=1}^{M-1}\sum_{i=1}^{N-1}(\delta y_{\bar{t}t})_i^j \varphi_i^j = \sum_{j=1}^{M-1}\sum_{i=1}^{N-1}(\delta \tilde{y}_{\bar{x}x})\varphi_i^j - \sum_{j=1}^{M-1}\sum_{i=1}^{N-1}(\hat{y}\delta p_i + p\delta\tilde{y}_i)\varphi_i^j \tag{44}$$

We will denote the items by $S_1, S_2, S_3$ consecutively in that expression and modify them separately:

$$S_1 = \frac{1}{\tau}\sum_{i=1}^{N-1}\sum_{j=1}^{M-1}\left(\frac{\delta y^{j-1} - \delta y^j}{\tau} - \frac{\delta y^j - \delta y^{j-1}}{\tau}\right)\varphi_i^j =$$

$$= \frac{1}{\tau^2}\sum_{i=1}^{N-1}[(\varphi, \Delta\delta y) - (\varphi, \nabla\delta y)]$$

Here
$$\Delta y^i = y^{j+1} - y^i, \quad \nabla y^j = y^{j-1}$$

$$(y,v) = \sum_{j=1}^{M-1} y^i v^j, \quad (y,v] = \sum_{j=1}^{M} y^j v^j$$

We apply to $S_1$ difference analogues of integration by parts:

$$(y, \Delta v) = -(v, \nabla y] + y^M v^M - y^0 v^1$$
$$(y, \nabla v) = -[v, \Delta y) + y^M v^{M-1} - y^0 v^0$$

Then the expression for $S_1$ will have the form:

$$S_1 = \frac{1}{\tau^2}\sum_{i=1}^{N-1}\left[\{-(\delta y, \nabla\varphi] + \delta y^M \varphi^M - \varphi^0 \delta y^1\} - \{-[\delta y, \Delta\varphi) + \varphi^M \delta y^{M-1} - \varphi^0 \delta y^0\}\right]$$

$$= \frac{1}{\tau^2}\sum_{i=1}^{N-1}\{[\delta y, \Delta\varphi) - (\delta y, \nabla\varphi] + \delta y^M \varphi^M - \varphi^0 \delta y^1 - \varphi^M \delta y^{M-1} + \varphi^0 \delta y_0\}$$

$$= \sum_{i=1}^{N-1}\left\{\frac{1}{\tau}\sum_{j=0}^{M-1}\delta y_i^j\left(\frac{\varphi_i^{j+1}-\varphi_i^j}{\tau}\right) - \frac{1}{\tau}\sum_{j=1}^{M}\delta y_i^j\left(\frac{\varphi^j-\varphi^{j-1}}{\tau}\right)\right\}$$

$$+ \delta y^M \varphi^M - \varphi^0 \delta y^1 - \varphi^M \delta y^{M-1} + \varphi^0 \delta y$$

Further (G.A.Tyulepberdinova), we bring to summing up over j from 1 to M – 1, we obtain:

$$S_1 = \sum_{i=1}^{N-1}\left[\frac{1}{\tau}\sum_{j=1}^{M-1}\frac{\varphi_i^{j+1}-\varphi_i^j}{\tau}\delta y_i^j - \frac{1}{\tau}\sum_{j=1}^{M-1}\frac{\varphi_i^j-\varphi_i^{j-1}}{\tau}\delta y_i^j + \delta y_i^0(\varphi_i^1-\varphi_i^0) - \right.$$

$$\left. - \delta y_i^M(\varphi_i^M-\varphi_i^{M-1}) + \delta y_i^M \varphi_i^M - \varphi_i^0\delta y_i^1 - \varphi_i^M \delta y_i^{M-1} + \varphi_i^0\delta_i^0\right]$$

After combining like terms and using the difference derivative for $\varphi_i^j$, we have:

$$S_1 = \sum_{i=1}^{N-1}\sum_{j=1}^{M-1}\delta y^j \varphi_{\bar{t}t} + \sum_{i=1}^{N-1}\left[\delta y^M \varphi^{M-1} - \varphi^M \delta y^M + \delta y^0 \varphi^1 - \varphi^0 \delta y^1\right]$$

Adding to the second term and then subtracting $\delta y_i^0 \varphi_i^0, \delta y_i^M \varphi_i^M$, we obtain

$$S_1 = \sum_{i=1}^{N-1}\sum_{j=1}^{M-1}\delta y^j \varphi_{\bar{t}t} + \sum_{i=1}^{N-1}\left[\delta y_i^0(\varphi_i^1-\varphi_i^0)\right.$$

$$\left. -\varphi_i^0(\delta y_i^1-\delta y_i^0) + \delta y_i^M(\varphi_i^{M-1}-\varphi_i^M) + \varphi_i^M(\delta y_i^M-\delta y_i^0)\right]$$

Using the condition (42) and assuming that $\varphi_i^M = 0, \varphi_{\bar{t}t}^M = 0$, we obtain

$$S_1 = \sum_{i=1}^{N-1}\sum_{j=1}^{M-1}\delta y^j \varphi_{\bar{t}t} \qquad (45)$$

We translate the expression for $S_2$. Expanding difference derivatives we have:

$$S_2 = \sum_{j=1}^{M-1}\frac{1}{h}\sum_{i=1}^{N-1}\left(\frac{\delta\tilde{y}_{i+1}-\delta\tilde{y}_i}{h} - \frac{\delta\tilde{y}_i-\delta\tilde{y}_{i-1}}{h}\right)\varphi_i^j$$

$$= \frac{1}{h^2}\sum_{j=1}^{M-1}\sum_{i=1}^{N-1}\left(\Delta\delta\tilde{y}_i\varphi_i^j - \nabla\delta\tilde{y}_i\varphi_i^j\right)$$

Using difference analogues of integration by parts, we have

$$S_2 = \frac{1}{h^2} \sum_{j=1}^{M-1} \left[ -(\hat{\delta y}_i, \nabla \varphi_i^j]_i + \varphi_N \delta y_N - \varphi_0 \delta y_1 + \right.$$
$$\left. + [\hat{\delta y}_i, \nabla \varphi_i^j) - \delta y_{N-1} \varphi_N + \varphi_0 \delta y_0 \right]$$

Expanding scalar products, we obtain:

$$S_2 = \frac{1}{h^2} \sum_{j=1}^{M-1} \left[ -\sum_{i=1}^{N} \hat{\delta y}_i, \nabla \varphi_i^j + \sum_{i=0}^{N-1} \hat{\delta y}_i \Delta \varphi_i^j \right.$$
$$\left. = \varphi_N \delta y_N - \varphi_0 \delta y_1 - \delta y_{N-1} \varphi_N + \varphi_0 \delta y_0 \right]$$

Expanding the sums and combining them again, we obtain:

$$S_2 = \frac{1}{h^2} \sum_{j=1}^{M-1} \left[ -\sum_{i=1}^{N-1} \hat{\delta y}_i \left( \Delta \varphi_i^j - \nabla \varphi_i^j \right) + \varphi_N^j \hat{\delta y}_N \right.$$
$$\left. -\varphi_0^j \hat{\delta y}_1 - \hat{\delta y}_{N-1} \varphi_N^j + \varphi_0^j \hat{\delta y}_0 - \hat{\delta y}_N \left( \varphi_N^j - \varphi_{N-1}^j \right) + \hat{\delta y}_0 \left( \psi_1^j - \varphi_0^j \right) \right]$$
$$= \frac{1}{h^2} \sum_{j=1}^{M-1} \left[ -\sum_{i=1}^{N-1} \hat{\delta y}_i \Delta \varphi_i^j - \nabla \varphi_i^j) - \varphi_0^j \hat{\delta y}_1 - \hat{\delta y}_{N-1} \varphi_N^j + \varphi_{N-1}^j \hat{\delta y}_N + \hat{\delta y}_0 \varphi_1^j \right]$$

Substituting $j' = j+1$, we have

$$S_2 = \frac{1}{h^2} \sum_{j'=2}^{M} \left[ -\sum_{i=1}^{N-1} \delta y_i^{j'} \left( \Delta \breve{\varphi}_i - \nabla \breve{\varphi}_i \right) + \right. \tag{46}$$

$$\left. + \breve{\varphi}_{N-1} \delta y_N^{j'} + \delta y_0^{j'} \breve{\varphi}_1 - \delta y_{N-1}^{j'} \breve{\varphi}_N - \breve{\varphi}_0 \delta y_1^{j'} \right] \tag{47}$$

Let us consider the last expression

$$S_3 = -\sum_{i=1}^{N-1} \sum_{j=1}^{M-1} \left( \hat{y} \delta p_i + p_i \hat{\delta y}_i \right) \varphi_i^j$$

If we substitute j' = j + 1 then

$$S_3 = -\sum_{i=1}^{N-1} \sum_{j'=2}^{M} y^{j'} \varphi_i^{j'-1} \delta p_i - \sum_{i=1}^{N-1} \sum_{j'=2}^{M-1} p_i \varphi_i^{j'-1} \delta y$$

Let us reduce it to the common summation over the index *j*. For this aim, let us consider the formula (45) for the expression for $S_1$. We have

$$S_1 = \sum_{i=1}^{N-1} \sum_{j=1}^{M-1} \delta y^j \varphi_{tt} + \sum_{i=1}^{N-1} \left[ \sum_{j=2}^{M-1} \delta y^j \varphi_{tt} + \delta y^1 \frac{1}{\tau} \left( \frac{\varphi_i^2 - \varphi_i^1}{\tau} - \frac{\varphi_i^1 - \varphi_i^0}{\tau} \right) \right] \tag{48}$$

Let us consider the following auxiliary dual problem:

$$\varphi_{tt} = \breve{\varphi}_{xx} - p_i \breve{\varphi}_i, \quad i = 1, 2, \ldots, N-1, \quad j = M-1, M-2, \ldots, 1, \quad (49)$$

$$\varphi_i^M = 0, \quad \varphi_{t,i}^M = 0, \quad i = 0, 1, 2, \ldots, N, \quad (50)$$

$$\varphi_N^j = 0, \quad j = M-2, M-1, \ldots, 0, \quad (51)$$

$$\varphi_{x,0}^j = 2[y_0^j - f_0^j], \quad j = M-2, M-1, \ldots, 0. \quad (52)$$

If we use the conditions (42) and (43), and the problem (49) – (52), then the expression (44) will have the form

$$\sum_{i=1}^{N-1} \delta y^1 \frac{1}{\tau} \left( \frac{\varphi_i^2 - \varphi_i^1}{\tau} - \frac{\varphi_i^1 - \varphi_i^0}{\tau} \right) = 2 \sum_{j=2}^{M} [y_0^j - f_0^j] \delta y_0^j - \sum_{i=1}^{N-1} \sum_{j=2}^{M-1} y^j \varphi_i^{j-1} \delta p_i$$

Which implies that

$$\Delta J(p) = h \sum_{i=1}^{N-1} \delta p_i \sum_{j=2}^{M-1} y^j \varphi_i^{j-1} + \sum_{i=1}^{N-1} \delta p_i \left( \frac{\varphi_i^2 - \varphi_i^1}{\tau} - \frac{\varphi_i^1 - \varphi_i^0}{\tau} \right)$$

Then the gradient of the functional has the form:

$$\nabla J(p) = \tau \sum_{j=2}^{M-1} y_i \varphi_i^{j-1} + \frac{1}{\tau} \left( \frac{\varphi_i^2 - \varphi_i^1}{\tau} - \frac{\varphi_i^1 - \varphi_i^0}{\tau} \right).$$

### 5 Conclusion

In this paper, the method of quickest descent applied to a one-dimensional inverse acoustic problem at a discrete level is considered. The gradient of functional at a discrete level is obtained. The algorithm of solving a problem using the optimization method at a differential level for a one-dimensional inverse acoustic problem is described. For numerical results (Напке M., Neuhauer A., Scherzer O.) of the simplified formulation of the inverse problem, the gradients of functional in continuous and discrete cases were compared in this article. The improved estimates of the rates of convergence of gradient-based methods considered were obtained. They are very important for practice because they provide with the possibility to make input data errors consistent with the iteration number. The practical application of this proposed new method of deriving the gradient of functional for an Acoustics discrete problem is that it makes calculations more accurate. The theoretical importance of the proposed method is the development of the technique of deriving estimates and gradients of functional at a discrete level.